# Automated Segmentation of Hip and Thigh Muscles in Metal Artifact-Contaminated CT using Convolutional Neural Network-Enhanced Normalized Metal Artifact Reduction

Mitsuki Sakamoto*[a], Yuta Hiasa[a], Yoshito Otake[a], Masaki Takao[b], Yuki Suzuki[a], Nobuhiko Sugano[b], Yoshinobu Sato[a]

Mitsuki Sakamoto: sakamoto.mitsuki.si2@is.naist.jp

[a]Graduate School of Science and Technology, Nara Institute of Science and Technology, 8916-5, Takayamacho, Ikoma, Japan 630-0192;
[b]Graduate School of Medicine, Osaka University, 2-2 Yamadaoka, Suita, Osaka, Japan 565-0871

**ABSTRACT**

In total hip arthroplasty, analysis of postoperative medical images is important to evaluate surgical outcome. Since Computed Tomography (CT) is most prevalent modality in orthopedic surgery, we aimed at the analysis of CT image. In this work, we focus on the metal artifact in postoperative CT caused by the metallic implant, which reduces the accuracy of segmentation especially in the vicinity of the implant. Our goal was to develop an automated segmentation method of the bones and muscles in the postoperative CT images. We propose a method that combines Normalized Metal Artifact Reduction (NMAR), which is one of the state-of-the-art metal artifact reduction methods, and a Convolutional Neural Network-based segmentation using two U-net architectures. The first U-net refines the result of NMAR and the muscle segmentation is performed by the second U-net. We conducted experiments using simulated images of 20 patients and real images of three patients to evaluate the segmentation accuracy of 19 muscles. In simulation study, the proposed method showed statistically significant improvement ($p<0.05$) in the average symmetric surface distance (ASD) metric for 14 muscles out of 19 muscles and the average ASD of all muscles from $1.17 \pm 0.543$ mm (mean ± std over all patients) to $1.10 \pm 0.509$ mm over our previous method. The real image study using the manual trace of gluteus maximus and medius muscles showed ASD of $1.32 \pm 0.25$ mm. Our future work includes training of a network in an end-to-end manner for both the metal artifact reduction and muscle segmentation.

## 1. INTRODUCTION

In total hip arthroplasty (THA), patient-specific analysis is important to evaluate surgical outcome and create appropriate rehabilitation plans. Segmentation of the muscles from patients' CT images allows a quantitative analysis of muscle atrophy and patient-specific biomechanical simulation. In medical image analysis, segmentation task has been studied extensively, and recently, remarkable success has been shown by deep learning based methods. For instance, U-net [1] is an architecture of the convolutional neural network (CNN) for biomedical image segmentation that is widely used since it is capable to achieve high performance with a limited number of training data. However, the network trained with preoperative CT images without containing metallic implant is not capable to achieve decent segmentation accuracy on the postoperative CT image that is contaminated by metallic artifact. Hence, we aim to develop an automated muscle segmentation method from postoperative CT images by combining a metal artifact reduction method and a CNN-based muscle segmentation method.

    Metal artifact reduction has been studied for a few decades, and Normalized Metal Artifact Reduction (NMAR) [2] is most popular and considered as one of the state-of-the-art methods. In NMAR, the projections within the metal trace are replaced with normalization and interpolation using a prior image. In this paper, we combined NMAR and U-net to develop the automated muscle segmentation from a postoperative CT image contaminated by metallic artifact.

This work extends preliminary versions of this work presented at IFMIA2019 [3]. We build on that work by employing a newly proposed U-net for the refinement of the NMAR results in addition to the U-net for muscle segmentation. In our preliminary work, NMAR results were input to the U-net for muscle segmentation without any refinement, thus the insufficiency of metal artifact reduction degraded the segmentation accuracy. Hence, in this work, we added the refinement network proposed by Gjesteby et al. [4] into our segmentation pipeline.

## 2. METHODS

### 2.1 Dataset

The database of 30 pre-operative CTs without metal artifact was constructed for the training of NMAR refinement network. These volumes were scanned in the axial plane for diagnosis of the patients subjected to THA surgery. The field of view was $360 \times 360$ mm$^2$, and the matrix size was $512 \times 512$. The slice thickness was 2.0 mm for the region including the pelvis and proximal femur, 6.0 mm for the femoral shaft region, and 1.0 mm for the distal femur region. A linear interpolation was performed to make the slice thickness of 2.0 mm uniform throughout the entire volume. In order to perform the simulation study, we simulated metal artifacts on these 30 CT data.

A separate database of 20 pre-operative CTs without metal artifact and three post-operative CTs with metal artifact was constructed for the training and evaluation of the muscle segmentation network. Nineteen muscle structures in hip and thigh regions and three bones (pelvis, femur and sacrum) were manually delineated using 20 pre-operative CTs by an expert surgeon. From three post-operative CTs, two muscles (gluteus maximus and medius muscle) was manually traced using images after NMAR by a computer scientist and verified by an orthopedic surgeon. Note that these two muscles were manually traced on the images after NMAR. The scanning protocol was the same as that of the other dataset.

### 2.2 Simulation of metal artifact

Metal artifact simulation was performed to create a dataset consisting of pairs of artifact free and artifact contaminated CT images. For the simulation, we first generated the region occupied by the metal object in a preoperative (artifact free) image by extracting metal regions from the real postoperative CT images by thresholding at 2,000 Hounsfield Unit (HU). The metals were manually aligned to a realistic location in the hip joint in the 20 preoperative CT images. For the remaining 10 preoperative CT images, we embedded the implant shape of a model provided by a manufacturer, so-called CAD model, in the pre-operative CTs to the location derived from the surgical planning data created by the operating surgeon.

Our metal artifact simulation pipeline basically follows the method described by Zhang et al. [5], except for the additional water beam-hardening correction method [6] to linearize the polychromatic x-ray projection. The flowchart of the metal artifact simulation pipeline used in our work is shown in Fig. 1. Firstly, for each voxel, the "*water weight*" and "*bone weight*", which represent the proportion of water equivalent tissue (e.g., skin, fat, organs) and bone tissue contained in each voxel, were calculated based on the intensity in the preoperative CT images by a method used in [7]. Following the original literature, we used a function that linearly transitions between water and bone at 100 and 1500 HU as shown in Fig. 1. Secondly, the *water-weighted* and *bone-weighted images* and another image where the metal region was filled with the linear attenuation coefficient of the metal were forward projected. The attenuation coefficients were calculated based on an equivalent monochromatic energy, 40 keV in our default setting. For the polychromatic projection, we obtained mass attenuation coefficients of the water, bone and iron from the XCOM dataset [8] and the x-ray spectrum of the incident ray using an online tool from Siemens [9] assuming the tungsten anode and peak tube voltage of 120 kVp (which was the setting used in the scanning of the images in our database). The spectrum between 1 keV and 120 keV was uniformly sampled at 1 keV interval as 120 monochromatic x-ray beams. In Fig. 1, as examples, the sinograms of the tissues projected with 40, 60 and 100 keV monochromatic x-rays are shown. Then, Poisson random noise was applied to the sinogram obtained based on the x-ray spectrum to simulate realistic quantum noise in the projection. Finally, after applying the water beam-hardening correction, the simulated image was obtained by the filtered-back projection using Hann filter. The entire process was implemented in Matlab (The Mathworks, Natic, MA). Our Matlab implementation can be found at https://github.com/NAIST-ICB/metal_artifact_simulation.

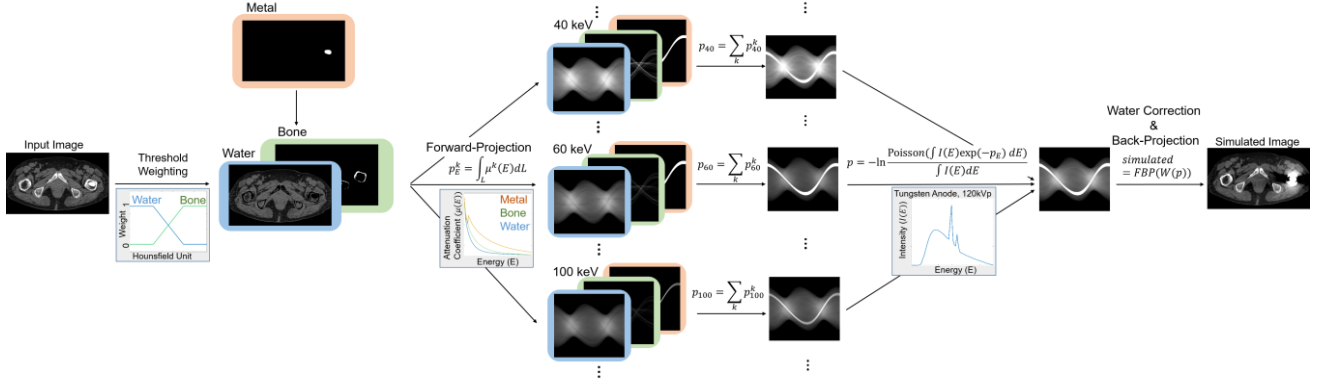

Fig. 1 Flowchart of the metal artifact simulation. The method basically follows Zhang et al. [5] except for the additional water correction [6] before the back projection step in order to reproduce the image similar to the one from common CT systems where a built-in calibration linearize the polychromatic x-ray projection.

### 2.3 Training of U-net

As a pre-processing for training of the refinement U-net, the image intensity was normalized by mapping [-150 350] HU to [0 255]. The refinement U-net was trained with Adam [10] optimizer, batch size of 2, 50 epochs and a learning rate of $1 \times 10^{-4}$. In the training of the refinement U-net, horizontal flip was applied to the training as data augmentation.

For training of muscle segmentation U-net, the same optimizer and parameters were set as above. A data augmentation for the segmentation network included horizontal flip, random rotation, shear transform, perspective skewing and random erasing [11] using a publicly available augmentation software Augmenter [12]. The U-net output often contains small island-like noise regions, thus, as a post-processing, we detected connected components and removed the components with a volume less than 5% of the total segmented volume.

For the evaluation of both networks, two-fold cross validation was performed. The training time was approximately 12 hours for each network on an Intel Xeon processor (3.2 GHz, 8 cores) with NVIDIA GeForce GTX 1080Ti. The average computation time for the inference on one CT volume with about 500 axial slices was approximately 2 minutes excluding file loading.

## 3. RESULTS

### 3.1 Simulation experiment

Fig. 2 shows quantitative evaluation results of the simulation experiment. The boxplots of the average symmetric surface distance (ASD) [13] of 19 muscles in 20 cases in the different settings are shown. The average and standard deviation of ASD values of all muscles from the original image, artifact-simulated image, images after artifact reduction by NMAR and by NMAR plus refinement U-net were 0.981 ± 0.465 mm, 1.64 ± 0.839 mm, 1.17 ± 0.543 mm and 1.10 ± 0.509 mm, respectively. The result shows statistically significant improvement by the refinement U-net in ASD on 14 out of 19 muscles ($p<0.05$) with Wilcoxon's signed rank test.

Fig. 3 illustrates a qualitative comparison among the original image, artifact-simulated image, images after artifact reduction by NMAR and by NMAR plus refinement U-net with 3D surface renderings of the segmentation results. The regions close to the implant in the axial slices at two levels are also shown below. Segmentation results were improved especially in the vicinity of the implant. For example, while U-net failed to segment gluteus maximus muscle (shown in yellow) in artifact-simulated image, a better segmentation accuracy was obtained in the artifact-reduced image by NMAR and refinement U-net. Additionally, the image quality improvement by the refinement U-net can be clearly observed from the axial slices.

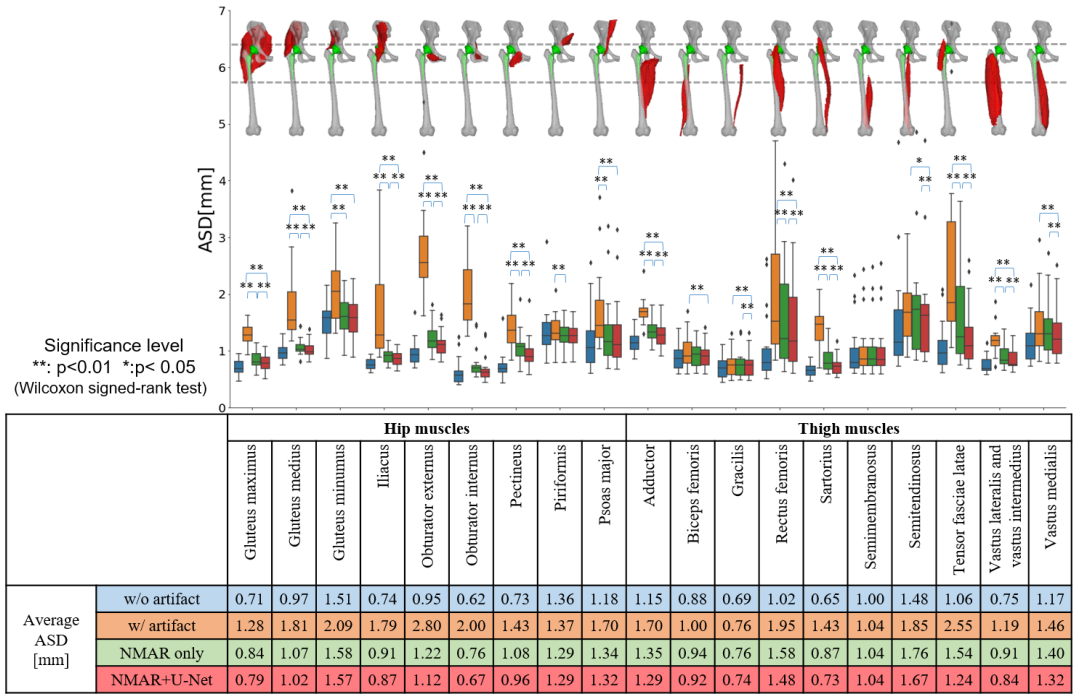

Fig. 2 Quantitative results of the simulation experiment. Boxplots showing ASD of 19 muscles of the original image (blue), artifact-simulated (orange), artifact-reduced by NMAR (green), and artifact-reduced by NMAR and refinement U-net (red). Average of ASD value over all patients is also shown numerically below the boxplots. The figure above each plot indicates the positional relationship between an implant and the target muscle (green shows the implant and the dotted lines indicate top and bottom edges of the implant).

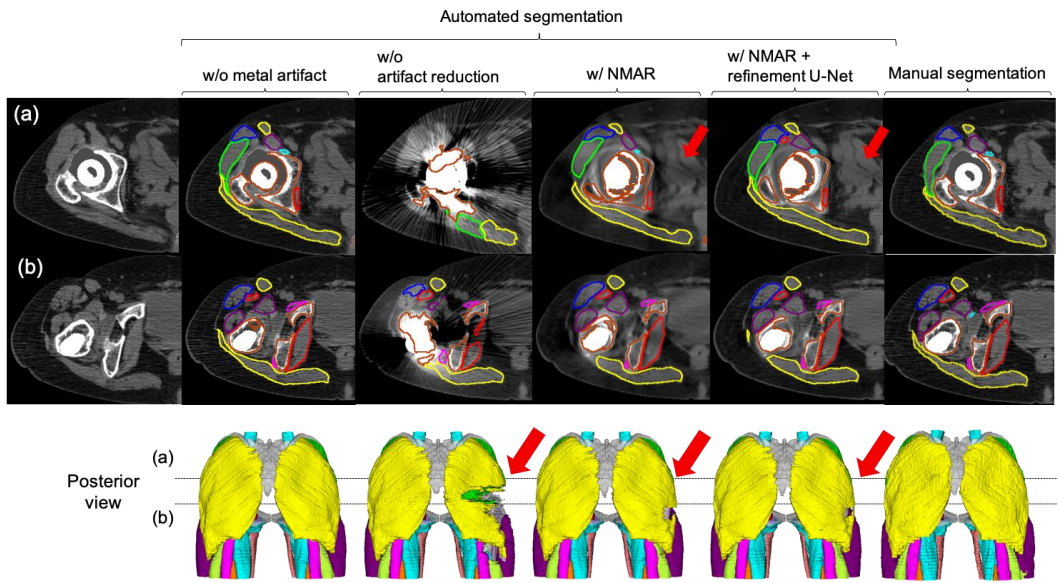

Fig. 3 A representative result of the simulation experiment. Axial cross section of the slices indicated by the dotted lines (upper) and 3D surface renderings of segmentation results from posterior views (lower) of a representative case. The window display in the cross section slices is [-150 350] HU. Red arrows indicate areas where accuracy improvements were observed when the method with NMAR and refinement U-net was used.

## 3.2 Real image experiment

We quantitatively evaluated the segmentation accuracy of the gluteus maximus and medius muscles from the real postoperative images that were contaminated by the metal artifact. ASD and Dice coefficient value of these two muscles are shown in Table 1. Fig. 4 shows 3D surface renderings of the segmented gluteus maximus (yellow), medius (green) muscles from the post-operative image, artifact-reduced image by NMAR and refinement U-net and the ground truth. Fig. 5 shows surface renderings of the 19 segmented muscles from anterior and posterior views, and axial cross sections of the postoperative images and the artifact corrected images by the proposed method, for qualitative evaluation. Note that the ground truth segmentation was not available for the muscles except for the two gluteus muscles shown in Fig. 4.

Table 1. Quantitative evaluation of the segmentation results based on ASD and DC.

|  |  | Gluteus maximus | | Gluteus medius | |
|---|---|---|---|---|---|
|  |  | w/o artifact reduction | NMAR+ refinement U-net | w/o artifact reduction | NMAR+ refinement U-net |
| Average ASD [mm] | Patient 1 | 1.66 | 1.55 | 1.49 | 1.18 |
|  | Patient 2 | 1.46 | 1.14 | 1.19 | 1.05 |
|  | Patient 3 | 1.69 | 1.23 | 5.50 | 1.75 |
| Average DC [%] | Patient 1 | 89.7 | 90.4 | 86.6 | 87.9 |
|  | Patient 2 | 90.7 | 92.6 | 91.1 | 91.7 |
|  | Patient 3 | 86.2 | 90.3 | 83.0 | 85.7 |

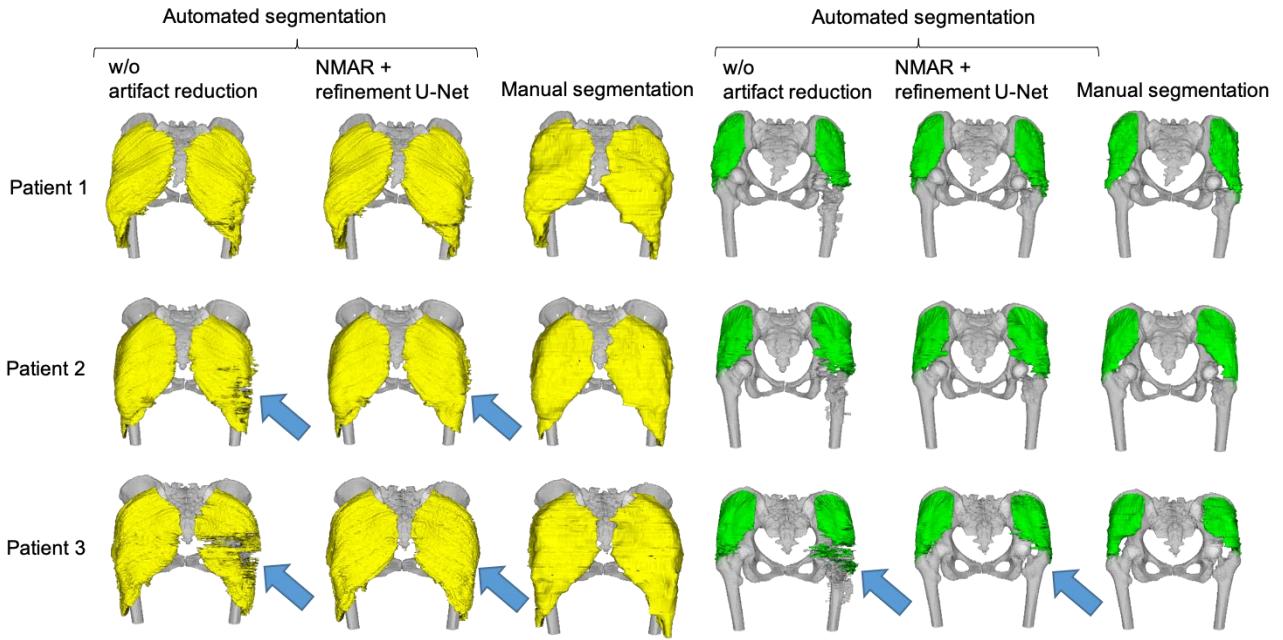

Fig. 4 Results of the real image experiment. 3D surface renderings of the ground truth and segmentation results of the gluteus maximus muscle (yellow) and medius muscle (green). Note that the ground truth for these muscles were created from images after NMAR. Arrows indicate areas where accuracy improvements were observed when the method with NMAR and refinement U-net was used.

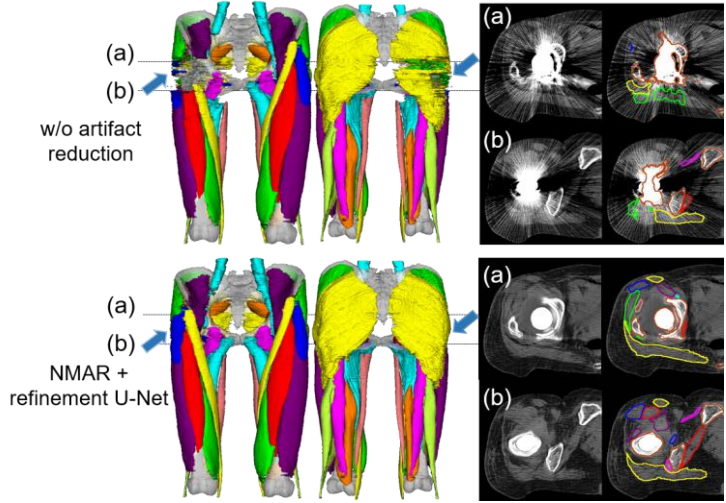

Fig. 5 A representative result of the real image experiment. 3D surface renderings of the segmentation results and axial cross sections for the real postoperative image of Patient 3 shown in Fig.4. The window display is [-150 350] HU. Note that the ground truth segmentation was not available for muscles except for the three gluteus muscles. Arrows indicate areas where accuracy improvements were observed when the method with NMAR and refinement U-net was used.

## 4. DISCUSSION AND CONCLUSION

In this paper, we proposed a segmentation method that combined NMAR and two U-nets consecutively to improve segmentation accuracy from the postoperative CT images of THA patients. Our results in the simulation experiment indicated that combining metal artifact reduction and segmentation methods is potentially effective in segmentation from postoperative image whose quality is severely degraded by the metal artifact.

One of the limitations of the proposed sequential approach that simply connects the artifact reduction and segmentation networks is that the segmentation result heavily depends on the output of the artifact reduction algorithm. For example, as shown in Fig. 5, the unclear boundaries of the musculoskeletal structure especially in the vicinity of the implant degraded the segmentation accuracy. In our method, the artifact reduction and the segmentation networks were trained individually. We consider that combining muscle segmentation and artifact reduction by an end-to-end training is one way to realize more effective artifact reduction and improve the accuracy of segmentation. Wu et al. [14] has proposed a method to detect abnormality in CT images by training a reconstruction network and an abnormality detection network in an end-to-end manner. They achieved a higher accuracy compared to the individually trained two networks.

Additionally, simulated images used in the training of the refinement network in our method highly affect the capacity of the refinement network. Although we took into account the physical effects such as beam-hardening caused by polychromatic projection and noise effect in metal artifact simulation, there is still difference between the simulated images and the real images, and the network is not able to obtain enough generalization capability due to this discrepancy. Generally, this kind of discrepancy is called domain shift and many previous works tackled this problem. Shrivastava et al. [15], for example, has proposed an adversarial learning method to reduce the gap between synthetic and real image distributions. We expect that the reduction of the discrepancy between the simulated and real images will lead to a higher segmentation accuracy in the real images.

Finally, our future work also includes application of the proposed method to clinical analysis such as assessment of the muscle volume, which is an important factor in the assessment of the outcome in total hip arthroplasty.

## ACKNOWLEDGEMENT

This work was partly supported by KAKENHI 19H01176 and 26108004.

## REFERENCES

[1] Ronneberger, O., Fischer, P., & Brox, T. (2015). U-net: Convolutional networks for biomedical image segmentation. *International Conference on Medical image computing and computer-assisted intervention*, 234-241.

[2] Meyer, E., Raupach, R., Lell, M., Schmidt, B., & Kachelrieß, M. (2010). Normalized metal artifact reduction (NMAR) in computed tomography. *Medical physics*, 37(10), 5482-5493.

[3] Sakamoto, M., Hiasa, Y., Otake, Y., Takao, M., Suzuki, Y., Sugano, N., & Sato, Y. (2019). Automated segmentation of hip and thigh muscles in metal artifact contaminated CT using CNN. In International Forum on Medical Imaging in Asia 2019. *International Society for Optics and Photonics*, 11050, 110500.

[4] Gjesteby, L., Shan, H., Yang, Q., Xi, Y., Claus, B., Jin, Y., et al. (2018). Deep Neural Network for CT Metal Artifact Reduction with a Perceptual Loss Function. *Proceedings of The Fifth International Conference on Image Formation in X-ray Computed Tomography*, 439-443.

[5] Zhang, Y., & Yu, H. (2018) Convolutional Neural Network based Metal Artifact Reduction in X-ray Computed Tomography. *IEEE transactions on medical imaging*, 37(6), 1370-1381.

[6] Herman, G. T. (1979). Correction for beam hardening in computed tomography. *Physics in Medicine & Biology*, 24(1), 81.

[7] Kyriakou, Y., Meyer, E., Prell, D., & Kachelrieß, M. (2010). Empirical beam hardening correction (EBHC) for CT. *Medical physics*, 37(10), 5179-5187.

[8] Berger, M., XCOM: photon cross sections database. http://www.nist.gov/pml/data/xcom/index.cfm. Accessed 21 June 2019.

[9] Simens Healthineers. Simulation of x-ray spectra. https://www.oem-xray-components.siemens.com/x-ray-spectra-simulation. Accessed 21 June 2019.

[10] Kingma, D. P., & Ba, J. (2014). Adam: A method for stochastic optimization. *arXiv preprint arXiv*:1412.6980.

[11] Zhong, Z., Zheng, L., Kang, G., Li, S., Yang, Y. (2017) Random erasing data augmentation. *arXiv preprint arXiv*:1708.04896.

[12] Bloice MD, Stocker C, Holzinger A. (2017), Augmentor: an image augmentation library for machine learning. *arXiv:*1708.04680.

[13] Van Ginneken, B., Heimann, T., & Styner, M., "3D segmentation in the clinic: A grand challenge," *3D segmentation in the clinic: a grand challenge*, 7-15 (2007).

[14] Wu, D., Kim, K., Dong, B., & Li, Q. (2017). End-to-end abnormality detection in medical imaging. *arXiv preprint arXiv*:1711.02074.

[15] Shrivastava, A., Pfister, T., Tuzel, O., Susskind, J., Wang, W., & Webb, R. (2017). Learning from simulated and unsupervised images through adversarial training. *In Proceedings of the IEEE Conference on Computer Vision and Pattern Recognition*, 2107-2116.